\documentstyle[preprint,aps]{revtex}
\newcommand{\be}{\begin{equation}} \newcommand{\ee}{\end{equation}} 
\newcommand{\bea}{\begin{eqnarray}}\newcommand{\eea}{\end{eqnarray}}

\begin{document}
\draft
\title{ Extended superconformal symmetry and Calogero-Marchioro model}
\author{Pijush K. Ghosh$^{*}$}
\address{ 
Department of Physics,
Ochanomizu University,\\
2-1-1 Ohtsuka, Bunkyo-ku,
Tokyo 112-8610, Japan.\\}
\footnotetext{$\mbox{}^*$E-mail address: 
pijush@degway.phys.ocha.ac.jp}  

\maketitle
\begin{abstract} 
We show that the two dimensional Calogero-Marchioro Model (CMM) without the
harmonic confinement can naturally be embedded into an extended $SU(1,1|2)$
superconformal Hamiltonian. We study the quantum evolution of the
superconformal Hamiltonian in terms of suitable compact operators of the
${\cal{N}}=2$ extended de Sitter superalgebra with central charge and discuss
the pattern of supersymmetry breaking. We also study the arbitrary $D$
dimensional CMM having dynamical $OSp(2|2)$ supersymmetry and point out the
relevance of this model in the context of the low energy effective action of
the dimensionally reduced Yang-Mills theory.
\end{abstract}
\narrowtext

\newpage

\section{introduction}

The Calogero-Moser-Sutherland (CMS) system is a class of
exactly solvable models in one dimension\cite{cs,cs1,pr,poly,sw}.
These models have been studied extensively from the
time of its inception more than thirty years ago and are well
understood. There are many higher dimensional generalizations
of these models \cite{cm,kr,mbs,pkg,fg,nn,mm,rsu}. Unfortunately, not a single
of these models are known to be exactly solvable or integrable. Among all
these systems,
the two dimensional Calogero-Marchioro model (CMM) deserves a special
attention for several reasons. First of all, for a certain value of the
coupling constant, different $n$-point correlation functions can be
calculated analytically by mapping this model to a complex Random Matrix
Theory\cite{kr,rsu,go}. This model also has been studied extensively 
\cite{rsu,kklz,imsc} in connection with several condensed matter systems like,
Quantum Hall effect, Quantum Dots, two dimensional Bose systems etc., revealing
many interesting features. 

The purpose of this paper is to unveil one more new feature of this model.
We first study the $D$ dimensional $N$-particle super-CMM with $N D$ bosonic and
$ N D$ fermionic degrees of freedom. We show how infinitely many exact
eigenstates can be constructed, both in supersymmetry-breaking
and supersymmetry-preserving phases, using the dynamical $OSp(2|2)$ symmetry
of the model.
We then show, within the specific formalism, only the two dimensional CMM
without the harmonic confinement can naturally be embedded into an extended
superconformal Hamiltonian. In other words, we construct an extended
${\cal{N}}=2$ superconformal version of the
two dimensional CMM. This construction is valid for arbitrary values of the
coupling constant and also for arbitrary $N$ number of particles. We study
the quantum evolution in terms of suitable compact operators of the extended
${\cal{N}}=2$ de Sitter superalgebra. Though we are able to find an infinite
number of exact eigenstates of these super-operators, the set is not complete
and we are unable to find the complete spectrum. We also discuss the
supersymmetry breaking pattern of the extended ${\cal{N}}=2$ de Sitter
supersymmetry with central charge and show how the half or the complete
breakdown of supersymmetry occurs. 
Finally, we point out the relevance of our findings in the context of
super-Yang-Mills theory.

We organize the paper in the following way. We first introduce the
conformal CMM model in arbitrary dimensions in the next section. An
infinite number of excited eigen states corresponding to the radial
excitations are constructed algebraically using the underlying $SU(1,1)$
symmetry. We construct the superconformal CMM in arbitrary dimensions in 
Sec..III. We also obtain infinitely many exact eigenstates using the
dynamical $OSp(2|2)$ symmetry of the model. The extended ${\cal{N}}=2$
superconformal CMM in $D=2$ is constructed in Sec. IV. The symmetry algebra
of the model and the supersymmetry-breaking pattern is discussed. Finally,
in Sec. V, we summarize
our findings and discuss the relevance of our results. We point out a
possible relation between the $D$ dimensional CMM considered in this
paper and the low energy effective action of the $D+1$ dimensional Yang-Mills
theory dimensionally reduced to $0+1$ dimension.

\section{conformal CMM}

We first consider the three operators $h$, ${\cal{D}}$ and $K$ given by,
\bea
&& h = \frac{1}{2} \sum_{i,\mu } p_{i,\mu }^2 + \frac{g}{2} ( g + D-2)
\sum_{i \neq j} \vec{r}_{ij}^{-2} + \frac{g^2}{2}  \sum_{i \neq j \neq k}
\left ( \vec{r}_{ij} . \vec{r}_{ik} \right ) \vec{r}_{ij}^{- 2}
\vec{r}_{ik}^{-2},\nonumber \\
&& {\cal{D}} = -\frac{1}{4} \sum_{i,\mu } \{ x_{i,\mu }, p_{i,\mu } \}, \ \
K = \frac{1}{2} \sum_{i,\mu } x_{i,\mu }^2, \ \
p_{i,\mu } = - i \frac{\partial}{\partial x_{i,\mu }}, \ \
\vec{r}_{ij} = \vec{r}_i - \vec{r}_j,
\label{eq0}
\eea
\noindent where $\vec{r}_i$ is the $D$ dimensional position vector of the $i$th
particle with $x_{i,\mu}$'s as the components and $g$ is the coupling constant.
We fix the convention that the
Roman indices run from $1$ to $N$, while the Greek indices run from $1$ to $D$.
These three operators admit the $O(2,1)$ algebra,
\be
[h, {\cal{D}}] = i h, \ \ [h, K] = 2 i {\cal{D}}, \ \ [K, {\cal{D}}] = - i K.
\label{eq1}
\ee
\noindent For the general conformal Hamiltonian, the many-body interaction
of $h$ (the last two terms ) should be replaced by a degree $-2$ homogeneous
function of the coordinates. The Hamiltonian $h$ describes the CMM without the
harmonic confinement. However, the ground-state of $h$ with the ground-state
energy $E=0$ is not even plane-wave normalizable. Following the prescription
suggested by de Alfaro, Fubini and Furlan \cite{dff} for such quantum
mechanical model with conformal symmetry, the quantum evolution can be
described by an appropriate compact operator. This compact operator can
be constructed from the linear combination of the Hamiltonian $h$, the
Dilatation generator ${\cal{D}}$ and the conformal generator $K$. Following
\cite{dff}, we choose this compact operator $H$ as, $H =  h + K$.
The introduction of $K$ breaks the scale invariance. The operator
$H$ is the $D$ dimensional CMM.

In one dimension, $H$ is exactly solvable and known as the
rational CMS Hamiltonian. In $D \geq 2$, though an infinitely many exact
eigenstates of this Hamiltonian can be found, the complete eigen-spectrum
is still not known. The ground-state wave-function is determined as
\cite{cm,kr},
\be
\psi_0= \prod_{i <j} {\mid \vec{r}_i - \vec{r}_j \mid}^g e^{-\frac{1}{2}
\sum_i \vec{r}_i^2},
\label{eq4}
\ee
\noindent with the ground-state energy $E_0=\frac{ND}{2} + g N (N-1)/2$.
Using the underlying $SU(1,1)$ symmetry,
\be
B_2^{\pm} = - \frac{1}{2} \left ( h - K \mp 2 i {\cal{D}} \right ),\ \
[H, B_2^{\pm}] = \pm 2 B_2^{\pm}, \ \ [ B_2^-,B_2^+] =  H,
\label{eq5}
\ee
\noindent one can construct infinitely many exact eigenstates of this
Hamiltonian. In particular,
\be
\psi_n = ( B_2^+ )^n \psi_0,
\label{ii}
\ee
\noindent are exact
eigenstates of $H$ with $E_n=E_0 + 2 n$. For $D=3$, these exact eigenstates
corresponding to the radial excitations were first obtained in \cite{cm}
by directly solving the Schr$\ddot{o}$dinger equation. Following the same
method,
these eigenstates were constructed for arbitrary $D$ in \cite{kr}. However,
we provide here an algebraic construction of these radial excitations in Eq.
(\ref{ii}), using the underlying $SU(1,1)$ symmetry. Unfortunately, the complete
spectrum of $H$ is still not known. The incompleteness of the spectrum can be
understood in the following way. In the limit $g \rightarrow 0$, the
Hamiltonian $H$ reduces to that of a system of $N$ free harmonic oscillators in 
$D$ dimensions. Thus, in this limit, the complete spectrum of a system
of $N$ free oscillators in $D$ dimensions should be reproduced. This is not the
case, as can be seen from the expressions $\psi_n$ and $E_n$ given above.

\section{ ${\cal{N}}=1$ superconformal CMM : $OSp(2|2)$}

We now construct the supersymmetric version of $h$ and $H$.
The supercharge $q$ and its conjugate $q^{\dagger}$ are defined as,
\be
q=\sum_{i,\mu} \psi_{i,\mu}^{\dagger} \ a_{i,\mu}, \ \ \ \ 
q^\dagger = \sum_{i,\mu} \psi_{i,\mu} \  a_{i,\mu}^{\dagger},
\label{eq6}
\ee
\noindent where the $N D$ fermionic variables $\psi_{i,\mu}$'s satisfy the
Clifford algebra,
\be
\{\psi_{i,\mu},\psi_{j,\nu}\}=0=\{\psi_{i,\mu}^{\dagger},\psi_{j,\nu}
^{\dagger}\}, \ \
\{\psi_{i,\mu}, \psi_{j,\nu}^{\dagger}\}=\delta_{ij} \delta_{\mu,\nu}.
\label{eq7}
\ee
\noindent The operators $a_i(a_i^{\dagger})$'s are analogous to bosonic 
annihilation ( creation ) operators. They are defined in terms of the
momentum operators
$p_{i,\mu}$ and the superpotential
$W(x_{1,1}, x_{1,2}, \dots, x_{1,D}, x_{2,1} \dots, x_{N,D-1}, x_{N,D})$ as,
\be
a_{i,\mu} = p_{i,\mu} - i W_{i\mu}, \ \ a_{i,\mu}^{\dagger}= p_{i,\mu} +
i W_{i,\mu},\
W_{i,\mu} = \frac{\partial W}{\partial x_{i,\mu}}.
\label{eq8}
\ee
\noindent For the general superconformal quantum mechanics, the superpotential
should have the following form,
\be
W = - ln G, \ \ \sum_{i, \mu} x_{i, \mu} \frac{\partial G}{\partial x_{i,\mu}}
= d \ G,
\label{eq8.1}
\ee
\noindent where $d$ is any arbitrary constant. We choose the superpotential
$W$ as,
\be
G = \prod_{i <j} {\mid \vec{r}_{ij} \mid}^g,
\label{eq9}
\ee
\noindent which results in the following Hamiltonian,
\bea
h_s & = & \frac{1}{2} \{q,q^{\dagger}\}\nonumber \\
& = & h + g \sum_{i \neq j; \mu} \left ( 2 \left ( x_{i,\mu} -
x_{j,\mu} \right )^2 \vec{r}_{ij}^{-2} - 1 \right ) 
\vec{r}_{ij}^{-2} \left (\psi_{i,\mu}^{\dagger} \psi_{i,\mu} -
\psi_{i,\mu}^{\dagger} \psi_{j,\mu} \right )\nonumber \\
&& +\ 2 g \sum_{i \neq j; \mu \neq \nu} \left ( x_{i,\mu} -
x_{j,\mu} \right )
\left ( x_{i,\nu} - x_{j,\nu} \right ) \vec{r}_{ij}^{-4} \left (
\psi_{i,\mu}^{\dagger} \psi_{i,\nu} -
\psi_{i,\mu}^{\dagger} \psi_{j,\nu} \right ).
\label{eq10}
\eea
\noindent The super-Hamiltonian $h_s$ is the supersymmetric generalization
of $h$. This can be checked by projecting $h_s$ in the zero-fermion sector
( $\psi_{i, \mu} |0> = 0$ ) of the $2^{ D N}$ dimensional fermionic Fock space.

The super-Hamiltonian $h_s$ does not have a normalizable
ground-state. Following the standard procedure in the literature
\cite{dff,fr,ap}, the quantum evolution can be described by the 
operator $R$ or $H_s$ defined as,
\be
H_s =  R + B - c, \ \ R = h_s + K, \ \
B = \frac{1}{2} \sum_{i,\mu} \left [\psi_{i,\mu}^{\dagger},
\psi_{i,\mu} \right ], \ \ c=\frac{g}{2} N (N-1) .
\label{eq11}
\ee
\noindent The new operator $H_s$ is the supersymmetric generalization of the
$D$ dimensional CMM $H$. The complete eigen-spectrum of this operator is known
\cite{me,fm} for $D=1$, both in supersymmetry preserving ( $g > 0$ ) as
well as supersymmetry breaking ( $g < 0$ ) phases. No attempt has been made
so far to study $H_s$ with its full generality for $D \geq 2$. We find that the ground-state of $H_s$ in the supersymmetric phase ($g > 0$) is determined as,
$\psi_s^0 = \psi_0 |0>$. A comment is in order at this point. The ground-state
wave-function $\psi_s^0$ is normalizable for $g > - \frac{1}{2}$. However, a
stronger criteria that each momentum operator $p_{i,\mu}$ is self-adjoint for
the wave-functions of the form $\psi_s^0$ requires $g > 0$. The supersymmetry
is preserved for $g > 0$, while it is broken for $g < 0$ \cite{fr,fm}. Let us
now define the following operators,
\bea
&& Q_1 = q - i S, \ \ Q_2 = q^{\dagger}- i S^{\dagger}, \ \
S = \sum_{i,\mu} {\psi_{i,\mu}}^{\dagger} x_{i,\mu},\nonumber \\
&& Q_1^{\dagger} = q^{\dagger} + i S^{\dagger}, \ \ 
Q_2^{\dagger}= q + i S, \ \
S^{\dagger} = \sum_{i,\mu} \psi_{i,\mu} x_{i,\mu}.
\label{eq13}
\eea
\noindent Note that the super-Hamiltonian $H_s = \frac{1}{2} \{Q_1,
Q_1^{\dagger} \}$. One can define bosonic and fermionic creation operators
\cite{fr,fm},
\be
{\cal{B}}_2^{\dagger} = - \frac{1}{4} \{ Q_1^{\dagger}, Q_2^{\dagger} \},
\ \ {\cal{F}}_2^{\dagger} = Q_2^{\dagger}.
\label{eq14}
\ee
\noindent It can be checked easily,
\be
[H_s, {\cal{B}}_2^{\dagger}] = 2 {\cal{B}}_2^{\dagger}, \ \
[H_s, {\cal{F}}_2^{\dagger}] = 2 {\cal{F}}_2^{\dagger}.
\label{eq15}
\ee
\noindent We construct a set of exact eigenstates with the help
of these operators. In particular,
\be
\psi_{n,\nu}= {\cal{B}}_2^{{\dagger}^n} {\cal{F}}_2^{{\dagger}^{\nu}}
\psi_s^0,
\label{eq16}
\ee
\noindent are the exact eigenstates of $H_s$ with the energy $E_{n,\nu}=
2 ( n + \nu ) $. The bosonic quantum number $n$ can take any non-negative
integer values, while the fermionic quantum number $\nu=0, 1$.
The super-Hamiltonian $H_s$ reduces to that of $N$ free
super-oscillators in $D$ dimensions in the limit $g \rightarrow 0$. In the
same limit, one would thus expect to obtain the complete eigen-spectrum of $N$
free super-oscillators in $D$ dimensions from $\psi_{n,\nu}$ and $E_{n,\nu}$.
Unfortunately, $E_{n, \nu}$ and $\psi_{n,\nu}$ describe only a small part of
the complete spectrum of the free super-oscillator Hamiltonian. Thus, the set
of exact eigenstates (\ref{eq16}) is not complete and we are unable to find the
complete spectrum.

The supersymmetry breaking phase of $H_s$ is characterized by $g <0$. A set
of exact eigenstates in this phase can also be constructed by using a duality
property of
this Hamiltonian. Consider a dual-Hamiltonian $\tilde{H}_s$ constructed
in terms of $Q_2$ and $Q_2^{\dagger}$ as, $\tilde{H}_s = \frac{1}{2}
\{ Q_2, Q_2^{\dagger} \}$. This Hamiltonian can also be obtained from $H_s$
by making $ g \rightarrow - g$ and $\psi_{i, \mu} \leftrightarrow
\psi_{i,\mu}^{\dagger}$ \cite{fm}. We determine
the ground-state of $\tilde{H}_s$ in its own supersymmetric phase ( $g < 0$) as,
\be
\tilde{\psi}_0 = \prod_{i <j} {\mid \vec{r}_i - \vec{r}_j \mid}^{-g}
e^{-\frac{1}{2} \sum_i \vec{r}_i^2} |ND>, \ \
\psi_{i,\mu}^{\dagger} |ND> = 0. 
\label{eq17}
\ee
\noindent Note that $\tilde{H}_s$ is related to $H_s$ by the following
relation,
\be
H_s = {\tilde{H}}_s + B - 2 c.
\label{eq18}
\ee
\noindent Thus, $\tilde{\psi}_0$ is also an exact eigenstate of $H_s$ with
the ground-state energy $E_0 = B - 2 c$, which is positive definite for
$g < 0$. This is in fact the ground-state wave-function of $H_s$ in the
supersymmetry breaking phase. A comment is in order at this point. Usually,
there are no general methods to find eigenstates in supersymmetry-breaking
phase of a model. However, the duality symmetry of $H_s$ plays an important
role to understand the supersymmetry-breaking phase of the model.
Firstly, the wave-function $\tilde{\psi}_0$ is guaranteed to be the
ground-state of $H_s$ for $g < 0$, because of the relation (\ref{eq18}) and the
fact that $\tilde{\psi}_0$ is the ground state of the dual-Hamiltonian
$\tilde{H}_s$ in its own supersymmetry-preserving phase $g < 0$ . Further, an
algebraic construction of excited states of $H_s$ for $g < 0$ is possible
using the duality symmetry. In particular,
a  set of excited states can be obtained by
acting different powers of the bosonic creation operator 
$\tilde{\cal{B}}_2^{\dagger}$ and the fermionic creation operator
$\tilde{\cal{F}}_2^{\dagger}$ on $\tilde{\psi}_0$, where these operators
are obtained from (\ref{eq14}) by making $ g \rightarrow -g $ and
$ \psi_{i,\mu} \leftrightarrow \psi_{i,\mu}^{\dagger}$. In particular,
the eigenstates and the corresponding eigenvalues are,
\be
{\tilde{\psi}}_{n, \nu} = \tilde{\cal{B}}_2^{{\dagger}^n}
\tilde{\cal{F}}_2^{{
\dagger}^{\nu}} \tilde{\psi}_0, \ \
\tilde{E}_{n, \nu} = E_0 + 2 ( n + \nu).
\ee
This set of exact eigenstates is not again complete.

\section{${\cal{N}}=2$ superconformal CMM : $SU(1,1|2)$}

After the center of mass separation, the super-Hamiltonian $h_s$ for $D=2$ and
$N=2$ reduces to the model considered in \cite{fr}. This model has been shown
to have extended $SU(1,1|2)$ superconformal symmetry \cite{fr}.
We generalize the work of \cite{fr} for arbitrary two dimensional $N$ particle
systems and find the criteria for having $SU(1,1|2)$ superconformal symmetry
in the following. The superpotential (\ref{eq8.1}) with the further constraint,
\be
G=  f(z_1, z_2, \dots, z_N) \ g(z_1^*, z_2^*, \dots, z_N^*), \ \
\ \ z_k = x_{k,1} + i x_{k,2}, \ \
\ \ z_k^* = x_{k,1} - i x_{k,2}, \ \
\label{eq19.0}
\ee
\noindent always gives rise to ${\cal{N}}=2$ superconformal Hamiltonian. The
homogeneity condition on $G$ implies that the (anti-)holomorphic function $(g)f$
should also be homogeneous. Note that except for the
two dimensional CMM and a nearest-neighbor variant of this model \cite{nn},
none of the other two dimensional model \cite{mbs,pkg,fg} satisfies the above
criteria. Thus, the two dimensional CMM enjoys a special status over all other
models. We specialize to $D=2$ and CMM in rest of the discussions.

Let us define an operator $Y$ and its conjugate $Y^{\dagger}$ as,
\be
Y = \frac{1}{2} \sum_i \epsilon_{\mu \nu} \psi_{i,\mu}
\psi_{i,\nu}, \ \
Y^{\dagger} = - \frac{1}{2} \sum_i \epsilon_{\mu \nu}
\psi_{i,\mu}^{\dagger} \psi_{i,\nu}^{\dagger}, \
\label{eq19.1}
\ee
\noindent where $\epsilon_{\mu \nu}$ is the two dimensional Levi-Civita
pseudo-tensor. We follow the convention that the repeated indices of the
Levi-Civita pseudo-tensor are always summed over. The operators $Y$,
$Y^{\dagger}$ and $B$ constitute a $SU(2)$ algebra,
\be
[Y, Y^{\dagger}] = - B, \ \ [B, Y] = - 2 Y, \ \ [ B, Y^{\dagger}] = 
2 Y^{\dagger}.
\label{eq19.2}
\ee
\noindent Further, we have the following commutation relations,
\be
\left [ Y^{\dagger}, \psi_{i,\mu} \right ] = 
\epsilon_{\mu \nu} \psi_{i, \nu}^{\dagger}= \bar{\psi}_{i, \mu}, \ \
\left [ Y, \psi_{i,\mu}^{\dagger} \right ] = -
\epsilon_{\mu \nu} \psi_{i, \nu} = - \bar{\psi}_{i,\mu}^{\dagger}. \ \
\label{eq19.3}
\ee
\noindent Following \cite{fr}, it can be shown that the unitary transformation
$U$, which represents a $180^{0}$ rotation around the second axis in the
internal space, performs the following transformation,
\be
U^{-1} \psi_{i,\mu} U = \bar{\psi}_{i, \mu}, \ \
U^{-1} \psi_{i,\mu}^{\dagger} U = \bar{\psi}_{i, \mu}^{\dagger}.
\label{eq19.4}
\ee
\noindent The $SU(2)$ generators $Y$, $Y^{\dagger}$ and $B$ commute with
the Hamiltonian $h_s$. The Hamiltonian $h_s$ has the internal $SU(2)$ symmetry
and is invariant under the unitary transformation $U$.

The extended ${\cal{N}}=2$ supersymmetry can be constructed by combining
together the $SU(2)$ generators, the operators $Q_1$, $Q_2$, $S$ and their
conjugates and a set of new operator $\bar{A}= U^{-1} A U $ corresponding
to each odd operator $A$. Define the new supercharges $\bar{q}$ and
$\bar{q}^{\dagger}$ following this prescription as\cite{fr},
\be
\bar{q} = \sum_{i,\mu} \bar{\psi}_{i,\mu}^{\dagger} a_{i,\mu}
= \sum_i \epsilon_{\mu,\nu} \psi_{i,\nu} a_{i, \mu}, \ \
\bar{q}^{\dagger} = \sum_{i,\mu} \bar{\psi}_{i,\mu} a_{i,\mu}^{\dagger}
= \sum_i \epsilon_{\mu,\nu} \psi_{i,\nu}^{\dagger} a_{i, \mu}^{\dagger}.
\label{eq19.5}
\ee
\noindent These supercharges satisfy the following anti-commutation relations
\cite{fr},
\be
\frac{1}{2} \{q,q^{\dagger} \} = h_s, \ \
\frac{1}{2} \{\bar{q}, \bar{q}^{\dagger} \} = h_s.
\label{eq19.6}
\ee
\noindent All other anticommutators among themselves vanish. The
super-Hamiltonian now will have a quartet structure. However, as noted earlier,
$h_s$ does not have a normalizable ground-state. The quantum evolution can
be described by $ R = h_s +K$ or $H_s$. We now explore the full $SU(1,1|2)$
symmetry. Define \cite{fr},
\bea 
&& \bar{Q}_1=\bar{q}-i \bar{S}, \ \bar{Q}_2= \bar{q}^{\dagger} -
i \bar{S}^{\dagger}, \ \bar{S}=\sum_{i,\mu} \bar{\psi}_{i,\mu}^{\dagger}
x_{i,\mu} = \sum_i \epsilon_{\mu \nu} \psi_{i, \nu} x_{i,\mu},\nonumber \\
&& \bar{Q}_1^{\dagger}=\bar{q}^{\dagger} + i \bar{S}^{\dagger}, \ 
\bar{Q}_2^{\dagger} = \bar{q} + i \bar{S}, \ 
\bar{S}^{\dagger} = \sum_{i,\mu} \bar{\psi}_{i,\mu}
x_{i,\mu} = \sum_i \epsilon_{\mu \nu} \psi_{i, \nu}^{\dagger} x_{i,\mu}.
\label{eq19.7}
\eea
\noindent The operators $Q_1$, $\bar{Q}_2$ and their conjugates have the
following anti-commutator algebra,
\bea
&& \frac{1}{2} \{ Q_1, Q_1^{\dagger} \} = R + B - c=H_s,\nonumber \\
&& \frac{1}{2} \{ \bar{Q}_2, \bar{Q}_2^{\dagger} \} = R + B + c = H_s +
2 c,\nonumber \\
&& \frac{1}{2} \{ Q_1, \bar{Q}_2^{\dagger} \} = - \frac{1}{2}
\{ Q_1^{\dagger}, \bar{Q}_2 \} = - i J,
\label{eq19.8}
\eea
\noindent where the angular momentum operator is defined as \cite{fr},
\be
J = \sum_i \epsilon_{\mu \nu} \left ( x_{i,\nu} p_{i, \mu} + i
\psi_{i, \mu}^{\dagger} \psi_{i,\nu} \right ).
\label{eq19.9}
\ee
\noindent Similarly the only non-vanishing anti-commutators among
$\bar{Q}_1$, $Q_2$ and their conjugates are,
\bea
&& \frac{1}{2} \{ Q_2, Q_2^{\dagger} \} = R - B + c = \tilde{H}_s,\nonumber \\
&& \frac{1}{2} \{ \bar{Q}_1, \bar{Q}_1^{\dagger} \} = R - B - c =
\tilde{H}_s - 2 c,\nonumber \\
&& \frac{1}{2} \{ Q_2, \bar{Q}_1^{\dagger} \} = - \frac{1}{2}
\{ Q_2^{\dagger}, \bar{Q}_1 \} = - i J.
\label{eq19.10}
\eea
\noindent All other non-vanishing anti-commutators are given by,
\bea
- \frac{1}{2} \{ Q_1, \bar{Q}_1^{\dagger} \} = \frac{1}{2}
\{ \bar{Q}_2, Q_2^{\dagger} \} = 2 Y^{\dagger}, \ \
- \frac{1}{2} \{ \bar{Q}_1, Q_1^{\dagger} \} = \frac{1}{2}
\{ Q_2, \bar{Q}_2^{\dagger} \} = 2 Y,\nonumber \\
\frac{1}{4} \{ Q_1, Q_2 \} = \frac{1}{4} \{ \bar{Q}_1, \bar{Q}_2 \}
= - {\cal{B}}_2, \ \
\frac{1}{4} \{ Q_1^{\dagger}, Q_2^{\dagger} \} =
\frac{1}{4} \{ \bar{Q}_1^{\dagger}, \bar{Q}_2^{\dagger} \}
= - {\cal{B}}_2^{\dagger}. \ \
\label{eq19.11}
\eea
\noindent The evolution can be described either by $H_s$ or $\tilde{H}_s$.

The supercharges $Q_1$ and $\bar{Q}_2$ are the generators of an extended
${\cal{N}}=2$ de Sitter supersymmetry with the central charge $c$. It is
amusing to note that the central charge $c$ is precisely the energy of
the classical minimum equilibrium configurations of the bosonic part of
$H_s$. However, we do not find any topological origin of $c$, as in the
case of field theories admitting soliton solutions in the
Bogomol'nyi-Prasad- Sommerfeld limit. As mentioned earlier,
$\psi_s^0$ is the ground state of $H_s$ in the supersymmetric phase. This
essentially implies that the supersymmetry associated with the generator
$\bar{Q}_2$ has broken. Thus, this is the case corresponding to
the spontaneous breakdown of supersymmetry from 
${\cal{N}}=2 \rightarrow {\cal{N}}=1$. For $g < 0$, the supersymmetry 
spontaneously breaks down completely.  The eigen-spectrum of $H_s$ in
this supersymmetry breaking phase can be constructed 
from $\tilde{H}_s$. 

The anti-commutator algebra (\ref{eq19.8}) is not in diagonal form
because of the last equation. The eigenstates of $H_s$ correspond
to the angular momentum eigenvalue $j=0$. Following \cite{fr} exactly,
let us define,
\be
\mu = cos \theta \ Q_1 + i sin \theta \ \bar{Q}_2, \ \
\nu= i sin \theta \ Q_1 + cos \theta \ \bar{Q}_2, \ \ tan (2 \theta )
= j/c.
\ee
\noindent It can be checked easily that,
\be
\frac{1}{2} \{\mu, \mu^{\dagger} \} = R + B - \sqrt{c^2+j^2}, \
\frac{1}{2} \{\nu, \nu^{\dagger} \} = R + B + \sqrt{c^2+j^2}, \
\{\mu, \nu^{\dagger} \} = 0.
\ee
\noindent The condition that the supersymmetric ground-state is annihilated by
both $\mu$ and $\mu^{\dagger}$ gives,
\bea
\psi_0^s(j) & = & \prod_{i <j} \left ( z_i - z_j \right )^{g^{-}} \left (
{z}_i^* - z_j^* \right )^{g^{+}} e^{- \frac{1}{2} 
\sum_i z_i z_i^*} |0>,\nonumber \\
g^{\mp} & = & \frac{1}{N(N-1)} \left [ (j^2+c^2)^{\frac{1}{2}} \mp j \right ].
\label{ig}
\eea
\noindent Note that for $j=0$, $g^+=g^-=\frac{g}{2}$ and $\psi_0^s(j=0)$
reduces to $\psi_0^s$. The eigenstates in (\ref{ig}) carry an angular
momentum,
\be
j = \frac{1}{2} \left ( g_+ - g_- \right ) N ( N - 1).
\ee
\noindent Note that $j$ receives contribution only from the bosonic part of
$\psi_0^s(j)$. Rest of the analysis can be carried out in a straightforward way.
In particular, one can verify easily,
\be
H_s(j) = \frac{1}{2} \{ \mu, \mu^{\dagger} \}, \ \
[H_s(j), {\cal{B}}_2^{\dagger}] = 2 {\cal{B}}_2^{\dagger}, \ \
[H_s(j), {\cal{F}}_2^{\dagger}] = 2 {\cal{F}}_2^{\dagger}.
\ee
\noindent Thus, we construct the excited states as,
\be
\psi_{n, \nu}(j) = {{\cal{B}}_2^{\dagger}}^n {{\cal{F}}_2^{\dagger}}^{\nu}
\psi_0^s(j),
\ee
\noindent where the bosonic quantum number $n$ can take any non-negative
integer values, while the fermionic quantum number $\nu=0, 1$. Note that
all these eigenstates have the same angular momentum.

\section{summary $\&$ discussions}

We have constructed and studied the $D$ dimensional superconformal 
CMM having dynamical $OSp(2|2)$ symmetry. Though we have obtained an infinite
number of exact states corresponding to the bosonic and the fermionic
excitations, the complete spectrum is still not known. Further, we have shown
that the two dimensional CMM can naturally be embedded into an extended
$SU(1,1|2)$ superconformal Hamiltonian. This construction of extended
${\cal{N}}=2$ superconformal many-particle Hamiltonian is valid for arbitrary
number of particles and also for arbitrary values of the coupling constant.
This is the central result of our paper. We have also studied the evolution
of this system in terms of operators of the extended ${\cal{N}}=2$ de Sitter
supersymmetry and discussed the supersymmetry-breaking pattern.

It may be worth mentioning here, recently, attempt to construct one
dimensional CMS Hamiltonian with extended superconformal symmetry has been
made \cite{nw}. It is found that within the specific formalism, the $SU(1,1|2)$
superconformal CMS model in one dimension can be constructed only for a certain
value of the coupling constant. Further, though a general formulation
of the multidimensional supersymmetric quantum mechanics with ${\cal{N}}=2$ was
given in \cite{rsu1}, no nontrivial many-particle systems of CMS-type have been
shown yet to result from such formulation. To the best of our knowledge, we are
not aware of any other work discussing $SU(1,1|2)$ superconformal Hamiltonian
of CMM-type with its full generality. Within this background, the extended
${\cal{N}}=2$ superconformal CMM presented in this letter appears to be the
first such example in the literature. The space-time dimensionality plays
an obvious role in our analysis. However, we would like to stress it again
that only the CMM and a nearest-neighbour variant of this model\cite{nn},
among several other interesting many-particle two dimensional
models\cite{mbs,pkg,fg}, are amenable for such a construction.

The history of studying supersymmetric quantum mechanical model with higher
number of supercharges\cite{cr} is long. One
of the major reason for the renewed interest in the (Super-)conformal Quantum
Mechanics is its relevance in the study of adS/CFT correspondence and
black holes\cite{bh}. Though a direct connection between the CMM and the black
hole physics can not be established at this point, we observe a possible
relation between the $D$ dimensional CMM and the low energy effective action
of $D+1$ dimensional Yang-Mills theory dimensionally reduced to $0+1$ dimension.
This observation is based on the existing results on this subject in the
literature\cite{rsu,mm}.

It is known\cite{rsu,kr,go} that the Hamiltonian $h$
for $D=2$ and $g=\frac{1}{2}$ describes the dynamics of a Gaussian ensemble
of $N \times N$ normal matrices in the limit $N \rightarrow \infty$. The
Gaussian action of the normal matrices is given by,
\be
{\cal{A}}(M, M^{\dagger}) = \frac{1}{4} \int dt \ Tr \left
( \frac{\partial M^{\dagger}}{\partial t}
\frac{\partial M}{\partial t} \right ), \ \
 \left [ M, M^{\dagger} \right ] = 0.
\label{d1}
\ee
\noindent The second equation defines $M$ to be normal matrices. The action
${\cal{A}}$ with $M$ as normal matrices is the low energy effective action of
$2+1$ dimensional Yang-Mills(YM) theory dimensionally reduced to $0+1$ dimension
with the choice of gauge $A_0=0$\cite{mm}. A term of the form
$[M,M^{\dagger}]^2$ drops out in the low energy limit giving rise to the
constraint on $M$ to be normal matrices. Thus, for the first time in the
literature, we observe the relation between the two
dimensional CMM with $g=\frac{1}{2}$ and the low energy effective action of
$2+1$ dimensional YM theory dimensionally reduced to $0+1$ dimension.
It is desirable to extend this result for arbitrary value of $g$,
much akin to the one dimensional CMS system.

It is worth recalling that an attempt to construct higher dimensional
generalizations of the
one dimensional CMS system from many-matrix models has been made in
\cite{mm}. At the classical level, the resulting Hamiltonian contains only
a two-body interaction term of the form $\sum_{i \neq j} \vec{r}_{ij}^{-2}$.
No trace of a three body term as in $h$ has been found. However,
for $D=2$, the many-matrix model considered in \cite{mm} is identical to
${\cal{A}}$ with $M$ as normal matrix which reduces to CMM with $g=\frac{1}{2}$
in the quantum mechanical treatment\cite{rsu}. Thus, it is expected that the
highly constrained classical models considered in \cite{mm}
should give rise to the CMM upon quantization for $D=2$. We also expect that
this will provide us a connection between the low energy effective action
of $2+1$ dimensional YM theory dimensionally reduced to $0+1$ dimension
and the two dimensional CMM for arbitrary value of $g$. Based on this
observation, we believe that the $D$ dimensional super-CMM considered in this
paper is in fact related to the low energy effective action of the $D+1$
dimensional super-YM theory dimensionally reduced to $0+1$ dimension. Since
the dimensionally reduced super-YM theory appears in many areas
of recent research activity like M-theory, D0-branes etc.\cite{qhs},
it is of immense
interest to put our belief relating CMM and super-YM on a firm footing.

\acknowledgements{ I would like to thank Tetsuo Deguchi for a careful reading
of the manuscript and valuable comments. This work is supported by a
fellowship (P99231) of the Japan Society for the Promotion of Science. I would
also like to acknowledge support in terms of a fellowship from the Institute
of Mathematical Sciences, Chennai, where a part of this work has been carried
out.}

\end{document}